\documentclass[compsoc, conference, a4paper, 10pt, times]{IEEEtran}

    \usepackage{cite}
    \usepackage{amsmath,amssymb,amsfonts}
    \usepackage{algorithmic}
    \usepackage{graphicx}
    \usepackage{textcomp}
    \usepackage{xcolor}
    \usepackage{booktabs}
    \usepackage{enumitem}
    \usepackage{hyperref}
    \usepackage{tabularx}

    \usepackage[acronym]{glossaries}
    \newacronym{suit}{SUIT}{Software Updates for Internet of Things}
\newacronym[plural=KEMs,firstplural=Key Encapsulation Mechanisms (KEMs)]{kem}{KEM}{Key Encapsulation Mechanism}
\newacronym[plural=DSAs,firstplural=Digital Signature Algorithms (DSAs)]{dsa}{DSA}{Digital Signature Algorithm (DSA)}
\newacronym{dhke}{DHKE}{Diffie-Hellman Key Exchange}
\newacronym{dh}{DH}{Diffie-Hellman}
\newacronym{ecdh}{ECDH}{Elliptic Curve Diffie-Hellman}
\newacronym{ecdsa}{ECDH}{Elliptic Curve Digital Signature Algorithm}
\newacronym{ecdhe}{ECDHE}{Elliptic Curve Diffie-Hellman Ephemeral}
\newacronym{hss}{HSS}{Hierarchical Signature System}
\newacronym{fts}{FTS}{Few-Time Signature (FTS)}
\newacronym{puf}{PUF}{Physical Unclonable Function}
\newacronym{ots}{OTS}{One-Time Signature}
\newacronym{lwe}{LWE}{Learning With Errors}
\newacronym{rot}{RoT}{Root of Trust}
\newacronym{svp}{SVP}{Shortest Vector Problem}
\newacronym{ec}{EC}{Elliptic Curve}
\newacronym{rsa}{RSA}{Rivest–Shamir–Adleman cryptosystem}
\newacronym{cps}{CPS}{Cyber-Physical System}

\newacronym[plural=CAs,firstplural=Certificate Authorities (CAs)]{ca}{CA}{Certificate Authority}
\newacronym{ra}{RA}{Registration Authority}
\newacronym{csr}{CSR}{Certificate Signing Request}
\newacronym[plural=DNs,firstplural=Distinguished Names (DNs)]{dn}{DN}{Distinguished Name}
\newacronym[plural=TAs,firstplural=Trust Anchors (TAs)]{ta}{TA}{Trust Anchor}
\newacronym{mac}{MAC}{Message Authentication Code}
\newacronym[plural=PKIs,firstplural=Public Key Infrastructures (PKIs)]{pki}{PKI}{Public Key Infrastructure}
\newacronym{api}{API}{Application Programming Interface}
\newacronym{mitm}{MitM}{Man-in-the-Middle}
\newacronym{cn}{CN}{Common Name}
\newacronym{san}{SAN}{Subject Alternative Names}
\newacronym{ebnf}{EBNF}{extended Backus-Naur form}
\newacronym{xml}{XML}{Extensible Markup Language}
\newacronym{xsl}{XSL}{Extensible Stylesheet Language}
\newacronym{pem}{PEM}{Privacy Enhanced Mail}
\newacronym{os}{OS}{Operating System}
\newacronym{jdk}{JDK}{Java Development Kit}
\newacronym{wget}{Wget}{GNU Wget}
\newacronym{curl}{cURL}{}
\newacronym{w3m}{W3M}{}
\newacronym{openssl}{OpenSSL}{}
\newacronym{gnutls}{GnuTLS}{}
\newacronym{gui}{GUI}{Graphical User Interface}
\newacronym{rng}{RNG}{Random Number Generator}
\newacronym{gtaapi}{GTA-API}{Generic Trust Anchor API}

\newacronym{ieee}{IEEE}{Institute of Electrical and Electronics Engineers}
\newacronym{ietf}{IETF}{Internet Engineering Task Force}
\newacronym{anima}{ANIMA}{Autonomic Networking Integrated Model and Approach}
\newacronym{aisec}{AISEC}{Fraunhofer Institute for Applied and Integrated Security}
\newacronym{tcg}{TCG}{Trusted Computing Group}
\newacronym{www}{WWW}{World Wide Web}
\newacronym{rfc}{RFC}{Request For Comments}
\newacronym{iec}{IEC}{International Electrotechnical Commission}
\newacronym{isa}{ISA}{International Society of Automation}
\newacronym{nist}{NIST}{National Institute of Standards and Technology}
\newacronym{bsi}{BSI}{Bundesamt für Sicherheit in der Informationstechnik}

\newacronym[plural=IDevIDs,firstplural=Initial Device Identifiers (IDevIDs)]{idevid}{IDevID}{Initial Device Identifier}
\newacronym[plural=LDevIDs,firstplural=Locally Significant Device Identifiers (LDevIDs)]{ldevid}{LDevID}{Locally Significant Device Identifier}
\newacronym[plural=DevIDs,firstplural=Device Identifiers (DevIDs)]{devid}{DevID}{Device Identifier}
\newacronym[plural=MASAs,firstplural=Manufacturer Authorized Signing Authorities (MASAs)]{masa}{MASA}{Manufacturer Authorized Signing Authority}
\newacronym[plural=JRCs,firstplural=Join Registrar and Coordinators (JRCs)]{jrc}{JRC}{Join Registrar and Coordinator}
\newacronym{acp}{ACP}{Autonomic Control Plane}
\newacronym{ani}{ANI}{Autonomic Networking Infrastructure}
\newacronym{prm}{PRM}{Pledge Responder Mode}
\newacronym{jws}{JWS}{JSON Web Signature}

\newacronym[plural=IACS,firstplural=Industrial Automation and Control Systems (IACS)]{iacs}{IACS}{Industrial Automation and Control System}
\newacronym[plural=SRs,firstplural=System Requirements (SRs)]{sr}{SR}{System Requirement}
\newacronym[plural=CRs,firstplural=Component Requirements (CRs)]{cr}{CR}{Component Requirement}
\newacronym[plural=REs,firstplural=Requirement Enhancements (REs)]{re}{RE}{Requirement Enhancement}
\newacronym[plural=SLs,firstplural=Security Levels (SLs)]{sl}{SL}{Security Level}
\newacronym[plural=SL-Ts,firstplural=Target Security Levels (SL-Ts)]{sl-t}{SL-T}{Target Security Level}
\newacronym[plural=SL-As,firstplural=Achieved Security Levels (SL-As)]{sl-a}{SL-A}{Achieved Security Level}
\newacronym[plural=SL-Cs,firstplural=Capability Security Levels (SL-Cs)]{sl-c}{SL-C}{Capability Security Level}
\newacronym[plural=PLCs,firstplural=Programmable Logic Controllers (PLCs)]{plc}{PLC}{Programmable Logic Controller}
\newacronym[plural=HMIs,firstplural=Human-Machine Interfaces (HMIs)]{hmi}{HMI}{Human-Machine Interface}
\newacronym[plural=MES,firstplural=Manufacturing Execution System (MESs)]{mes}{MES}{Manufacturing Execution System}
\newacronym[plural=ISPs,firstplural=Integration Service Providers (ISPs)]{isp}{ISP}{Integration Service Provider}
\newacronym[plural=MSPs,firstplural=Maintenance Service Providers (MSPs)]{msp}{MSP}{Maintentance Service Provider}

\newacronym[plural=SEs,firstplural=Secure Elements (SEs)]{se}{SE}{Secure Element}
\newacronym[plural=HTAs,firstplural=Hardware Trust Anchors (HTAs)]{hta}{HTA}{Hardware Trust Anchor}
\newacronym[plural=TPMs,firstplural=Trusted Platform Modules (TPMs)]{tpm}{TPM}{Trusted Platform Module}
\newacronym{iot}{IoT}{Internet of Things}
\newacronym{iiot}{IIoT}{Industrial Internet of Things}
\newacronym{wot}{WoT}{Web of Trust}
\newacronym{email}{e-mail}{Electronic Mail}
\newacronym{e2e}{E2E}{end-to-end}
\newacronym{ecc}{ECC}{Elliptic Curve Cryptography}
\newacronym{var}{VAR}{Value-Added Reseller}
\newacronym[plural=OIDs,firstplural=Object Identifiers (OIDs)]{oid}{OID}{Object Identifier}
\newacronym{uri}{URI}{Uniform Resource Identifier}
\newacronym{url}{URL}{Uniform Resource Locator}
\newacronym{mdm}{MDM}{Mobile Device Management}
\newacronym[plural=ASAs,firstplural=Autonomic Service Agents (ASAs)]{asa}{ASA}{Autonomic Service Agent}
\newacronym{json}{JSON}{JavaScript Object Notation}
\newacronym{tofu}{TOFU}{Trust On First Use}
\newacronym{it}{IT}{Information Technology}
\newacronym{ot}{OT}{Operational Technology}
\newacronym{scada}{SCADA}{Supervisory Control and Data Acquisition}
\newacronym{dos}{DoS}{Denial of Service}
\newacronym{dmz}{DMZ}{Demilitarized Zone}
\newacronym{cpu}{CPU}{Central Processing Unit}
\newacronym{ram}{RAM}{Random Access Memory}
\newacronym[plural=VMs,firstplural=Virtual Machines (VMs)]{vm}{VM}{Virtual Machine}
\newacronym{gpio}{GPIO}{General Purpose Input/Output}
\newacronym{spi}{SPI}{Serial Peripheral Interface}
\newacronym[plural=RPIs,firstplural=Raspberry Pis (RPIs)]{rpi}{RPI}{Raspberry Pi}
\newacronym[plural=VDIs,firstplural=Virtual Disk Images (VDIs)]{vdi}{VDI}{Virtual Disk Image}
\newacronym{ssd}{SSD}{Solid State Drive}
\newacronym{nat}{NAT}{Network Address Translation}
\newacronym{cli}{CLI}{Command-line Interface}
\newacronym{ui}{UI}{User Interface}
\newacronym[plural=OEMs,firstplural=Original Equipment Manufacturer (OEMs)]{oem}{OEM}{Original Equipment Manufacturer}
\newacronym{maas}{MaaS}{MASA-as-a-Service}
\newacronym{opcua}{OPC UA}{Open Platform Communications Unified Architecture}
\newacronym{ssi}{SSI}{Self Sovereign Identity}
\newacronym[plural=LLNs,firstplural=Low-Power and Lossy Networks (LLNs)]{lln}{LLN}{Low-Power and Lossy Network}
\newacronym{hids}{HIDS}{Host-based intrusion detection system}

\newacronym{ssl}{SSL}{Secure Sockets Layer}
\newacronym{vpn}{VPN}{Virtual Private Network}
\newacronym{tls}{TLS}{Transport Layer Security}
\newacronym{dtls}{DTLS}{Datagram Transport Layer Security}
\newacronym{ftp}{FTP}{File Transfer Protocol}
\newacronym{ldap}{LDAP}{Lightweight Directory Access Protocol}
\newacronym{imap}{IMAP}{Internet Message Access Protocol}
\newacronym{smtp}{SMTP}{Simple Mail Transfer Protocol}
\newacronym{sip}{SIP}{Session Initiation Protocol}
\newacronym{crl}{CRL}{Certificate Revocation List}
\newacronym{ocsp}{OCSP}{Online Certificate Status Protocol}
\newacronym{ip}{IP}{Internet Protocol}
\newacronym{tcp}{TCP}{Transmission Control Protocol}
\newacronym{http}{HTTP}{Hypertext Transfer Protocol}
\newacronym{https}{HTTPS}{Hypertext Transfer Protocol Secure}
\newacronym{html}{HTML}{Hypertext Markup Language}
\newacronym{xslt}{XSLT}{Extensible Stylesheet Language Transformation}
\newacronym{cmp}{CMP}{Certificate Management Protocol}
\newacronym{cmc}{CMC}{Certificate Management Protocol using the Cryptographic Message Syntax}
\newacronym{cms}{CMS}{Cryptographic Message Syntax}
\newacronym{scep}{SCEP}{Simple Certificate Enrolment Protocol}
\newacronym{est}{EST}{Enrollment over Secure Transport}
\newacronym{brski}{BRSKI}{Bootstrapping Remote Secure Key Infrastructure}
\newacronym{cbrski}{cBRSKI}{Constrained BRSKI}
\newacronym{rest}{REST}{Representational State Transfer}
\newacronym{rpl}{RPL}{IPv6 Routing Protocol for Low-Power and Lossy Networks}
\newacronym{ipsec}{IPsec}{Internet Protocol Security}
\newacronym{dhcp}{DHCP}{Dynamic Host Configuration Protocol}
\newacronym{lan}{LAN}{Local Area Network}
\newacronym{wan}{WAN}{Wide Area Network}
\newacronym{dns}{DNS}{Domain Name System}
\newacronym{ssh}{SSH}{Secure Shell}
\newacronym{ct}{CT}{Certificate Transparency}
\newacronym{coaps}{CoAPS}{Secure Constrained Application Protocol}
\newacronym{coap}{CoAP}{Constrained Application Protocol}
\newacronym{grasp}{GRASP}{Generic Autonomic Signaling Protocol}
\newacronym{sztp}{SZTP}{Secure Zero Touch Provisioning}
\newacronym{dpp}{DPP}{Device Provisioning Protocol}
\newacronym{pqc}{PQC}{Post-Quantum Cryptography}
\newacronym{pq}{PQ}{Post-Quantum}

\newacronym{ek}{EK}{Endorsement Key}
\newacronym{iak}{IAK}{Initial Attestation Key}
\newacronym{lak}{LAK}{Local Attestation Key}

\newglossaryentry{802.1ar}
{
    name={IEEE 802.1AR},
    description={An IEEE standard for secure device identifiers and device enrollment in industrial control systems}
}

\newglossaryentry{isaiec}
{
    name={IEC 62443},
    description={An international series of standards that address cybersecurity for operational technology in automation and control systems}
}

\newglossaryentry{pledge}
{
    name={pledge},
    plural={pledges},
    description={An device that wishes to bootstrap via BRSKI into a network}
}

\newacronym{gka}{GKA}{Group Key Agreement}
\newacronym{fs}{FS}{Forward Secrecy}
\newacronym{pcs}{PCS}{Post-compromise security}
\newacronym{can}{CAN}{Controller Area Network}
\newacronym{gracybus}{GRACYBUS}{GRoup key Agreement on Cyber-phYsical system BUSes}

\newacronym{mls}{MLS}{Messaging Layer Security}
\newacronym{kdf}{KDF}{Key Derivation Function}
    \usepackage[english]{babel}
    \addto\extrasenglish{%

    }

    \usepackage[all]{nowidow}
    \usepackage{longtable}

    \usepackage{array}
    \usepackage{cleveref}
    \usepackage{pifont}
    \usepackage{arydshln}
\begin{document}

\title{Secure Group Key Agreement on Cyber-Physical System Buses} 











\iftrue 

\author{
\IEEEauthorblockN{
Sebastian N. Peters\IEEEauthorrefmark{1}\IEEEauthorrefmark{2},
Lukas Lautenschlager\IEEEauthorrefmark{1}\IEEEauthorrefmark{2},
David Emeis\IEEEauthorrefmark{1}\IEEEauthorrefmark{2},
Jason Lochert\IEEEauthorrefmark{1}
}
\IEEEauthorblockA{\IEEEauthorrefmark{1}Technical University of Munich (TUM), Garching bei München, Germany}
\IEEEauthorblockA{\IEEEauthorrefmark{2}Fraunhofer AISEC, Garching bei München, Germany \\
\{sebastian.peters,lukas.lautenschlager\}@aisec.fraunhofer.de}
}
\fi

\maketitle


\begin{abstract}
\glspl{cps} rely on distributed embedded devices that often must communicate securely over buses.
Ensuring message integrity and authenticity on these buses typically requires group-shared keys for \glspl{mac}.
To avoid insecure fixed pre-shared keys and trust-on-first-use concepts, a \gls{gka} protocol is needed to dynamically agree on a key amongst the devices.
Yet existing \gls{gka} protocols lack adaptability to constrained \gls{cps} buses.

This paper targets authenticated, fully distributed \gls{gka} suitable for bus topologies under  constraints of industrial and cyber-physical systems, including broadcast-only links, half-duplex operation, resource limits, dynamic membership (including unannounced leaves), a long device lifetime, and a strong Dolev–Yao adversary capable of partitioning the bus.
We first systematise existing protocols, then derive the requirements necessary for an authenticated and fully distributed \gls{gka} on bus systems.
Finally, we design, implement, and evaluate a custom \gls{gka} protocol based on TreeKEM. 

\end{abstract}

\section{Introduction}
Modern industrial machinery and cyber-physical systems interconnect large numbers of embedded devices over broadcast buses such as \gls{can}, PROFIBUS, and Automotive and Industrial Ethernet. 
To ensure message integrity and authenticity on these buses, deployments typically use \glspl{mac}, which in turn require a securely shared group key. 
Establishing such a key without insecure fixed pre-shared keys or trust-on-first-use approaches necessitates a \gls{gka} protocol.

Conventional pairwise protocols, such as the \gls{tls} handshake, with certificate validation and derivation of session keys, work well for two-party communication but do not scale to groups. 
Each pairwise \gls{tls} on a bus results in exponential state and message complexity, multiplies authentication work and therefore conflicts with broadcast-oriented communication. 
Hence, group settings require a dedicated \gls{gka} protocol. 
Existing \glspl{gka} span a spectrum of designs: 
centralised schemes with a preselected leader that determines group membership and group keys; contributory schemes in which members collaboratively contribute entropy; and diverse internal organisations ranging from tree-based to ring-based or non-hierarchical structures. 
Some protocols assume static, always-online groups, whereas others support dynamic membership with runtime joins and leaves.

\gls{cps} buses introduce additional constraints that challenge existing \glspl{gka}. 
Devices are often resource-constrained due to cost and space budgets; subsets of devices may be powered on and off without warning; links may be broadcast-only or half-duplex; and deployments have long lifecycles that demand cryptographic agility. 
These realities limit the applicability of \gls{gka} designs originating from IT or IoT contexts, motivating a bus-centric approach.

Our research questions are:
\begin{enumerate}[label=\textbf{RQ\arabic*}, leftmargin=*, noitemsep]
\item What are the requirements for group key agreement between devices on a \gls{cps} bus?
\item Which existing \gls{gka} protocols would be suitable?
\item To what extent can our customised \gls{gka} protocol fulfil the requirements?
\end{enumerate}

Contributions. 
We define a bus-centric system model and derive 12 concrete requirements for secure and practical \glspl{gka} on \gls{cps} buses (\Cref{sec:req}). 
We then systematise existing \gls{gka} protocols and analyse them against these requirements, identifying TreeKEM as the most suitable foundation for our setting (\Cref{sec:protocol_analysis}). 
Building on TreeKEM, we design the \textbf{\gls{gracybus}} protocol to meet the stated requirements (\Cref{sec:protocol_design}). 
Finally, we evaluate its security and performance to demonstrate feasibility under resource constraints (\Cref{sec:evaluation}).
\Cref{sec:conclusion} concludes and outlines directions for future work.

\section{Background} \label{sec:background}
This section summarizes some foundational concepts necessary for \gls{gka} on buses. 
We first review bus architectures and their operational constraints that shape protocol requirements. 
Next, we introduce binary tree structure terminology that will be later used by tree-based \gls{gka}.

\subsection{Bus Systems}
Bus systems and, in particular, fieldbuses are still prevalent today and in the future due to their long lifecycles \cite{vukovic_digital_2021}. 
Those fieldbuses are still in use because they enable simple cabling, easy maintenance, and the simultaneous connection of multiple systems. They are therefore fitting for any digital system that has some constraints on manufacturing price or space. Other work also discusses that compatibility with existing fieldbus protocols is one major challenge in the field \gls{iiot} \cite{ungurean_software_2020}. This leads to the development of solutions looking to replace the fieldbus with different technology stacks \cite{trifonov_opc_2023}. Nevertheless, it is not expected that deployments will transition to modern technology stacks in the near future.

Common bus technologies include CAN \cite{ISO_11898} and PROFIBUS \cite{germancommissionforelectricalelectronicandinformationtechnologiesofdinandvde_Industrial_}. Those protocols are known for their limited properties. For example, \gls{can} has limited properties, such as half-duplex communication and a maximum payload size of 64 bytes \cite{ISO_11898}.
However, in recent years, the protection of messages transmitted via these protocols has become a requirement, which presents challenges for any security protocol added to these bus technologies.

This requirement for security motivates our work by presenting a protocol that enables agreement on a key in an authenticated and distributed manner. It has been developed with the limitations in mind, such as limited bandwidth and the requirement for dynamism for joining and leaving nodes. Our work focuses on group key agreements that align broadcast-only bus communication (e.g., all messages are received by anyone on the bus) with key establishment. This shared key can then be further used by the bus participants for encryption or integrity protection of messages.

\subsection{Binary Tree Operations}
Binary trees are a hierarchical data structure composed of \emph{nodes}. 
Each node can store application-specific data (e.g., primitive values or cryptographic material), and typically holds references to its left and right child; optionally, nodes may also hold a pointer to their parent. 
The single node at the top is the \emph{root}; nodes without children are \emph{leaves}.
The height of a binary tree is the number of nodes between the root and its deepest (furthest) leaf.
A \emph{subtree} is defined by choosing any node as a root and taking that node together with all of its descendants.
For any (leaf) node, its \emph{path} is the ordered set of nodes up to the tree’s root (inclusive). 
The \emph{co-path} is the set of the siblings of each node on the path, since the root has no siblings, the co-path has one fewer element than the path.

\section{System Model and Requirements} \label{sec:req}

This section describes our assumed generalised bus system.
We describe the necessary assumptions, detail an attacker model, and derive requirements from our use case.

\subsection{System Description and Assumptions}

The system model in our work is based on systems that are still widely in use. 
Those include, for example, cars and trains, which have a long lifespan and utilize bus protocols like \gls{can} because they are cost-effective and do not require additional networking equipment. 
Protocols like \gls{can} have limited properties, such as half-duplex communication and a maximum payload size of 64 bytes. \cite{ISO_11898}
However, in recent years, the protection of messages transmitted via these protocols has become a requirement, which presents challenges for any security protocol added to these bus technologies.

To make the key agreement protocol as agnostic as possible to the technology, we assume certain properties. 

- A Broadcast bus system on which multiple devices communicate
- Large enough payload sizes for cryptographic messages and entities (this excludes classic \gls{can} with only 8-byte payload)
- Authentication of devices is ensured via a common trust anchor, such as an industrial \gls{pki}
- Resource-constrained devices need to avoid excessive asymmetric operations; they are only acceptable once for an initial \gls{gka} or renewal
- no central entity in the network can be used to perform key distribution

Industrial \gls{pki} \cite{heinl_Standard_2023}...
For the initial deployment of trust anchors, protocols such as \gls{brski} have proven to function effectively even in industrial environments \cite{heinl_Leveraging_2025}.
First of all, we assume no central entity in the network can be used to perform key distribution. 
Therefore, we require a distributed key agreement protocol that allows any participant to perform the required operations. 
This aligns with the outlined use case in trains, where a split of different parts of the train is possible, therefore not allowing for a single central entity. 
For the same reason, our protocol designed in this work should allow dynamic group changes. 
Additionally, to avoid any dependency on the availability of a single participant, the protocol does not require one system that is always up.

\begin{figure}[t]
   \centering      
   \includegraphics[width=\columnwidth]{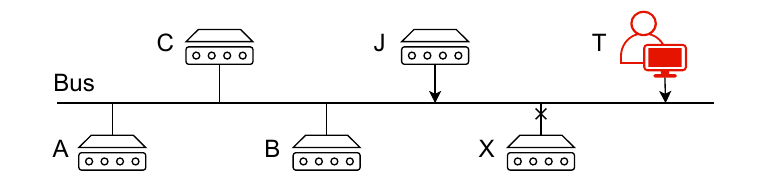}     
 \caption{System model: Bus system with connected devices A, B, and C. 
 Device J is joining the bus. Device X is leaving the bus abruptly. The attacker T has physical access to the bus.}
 \label{fig:sysmodel}
\end{figure}

\subsection{Attacker model}
\label{sec:att_model}

The underlying threat model we assume is the Dolev-Yao model~\cite {dolev_security_1983} to evaluate the security guarantees of the developed protocol. 
The Dolev-Yao model abstracts the communication channel and allows an attacker the following capabilities. 
An attacker can read, modify, replay, inject, drop, and reorder messages exchanged between participants on the bus. 
The model assumes that cryptographic primitives cannot be broken by an attacker. 

\subsection{Requirement derivation}

\begin{table*}[t]
  \centering
  \caption{Protocol Requirements for an ideal Secure \gls{gka} on Cyber-Physical System Buses}
  \label{tab:req}
  \begin{tabular}{@{}llp{0.7\linewidth}@{}}
    \toprule
    R\# & Name & Explanation \\
    \midrule
    R01 & Authenticity & Only authenticated participants are allowed to send valid protocol messages \\
    R02 & Fully Distributed & Every node can perform all protocol operations.  \\
    R03 & Independent & Operates without external services during normal use. \\
    R04 & Bus-Agnostic & Independent on any other protocol information (e.g., MAC address for ordering). \\
    R05 & Dynamic & Nodes can join and leave dynamically. Inactive nodes will be removed after a certain period, but this will not disrupt the normal protocol process. \\
    R06 & Scalable & Efficient for larger groups. Modern cyber-physical systems can contain over 100 devices. \\ 
    R07 & Dolev-Yao Attacker &  Robust to Dolev-Yao adversaries. \\
    R08 & \acrfull{pcs} & Compromisation of one node heals after a certain period of time. \\
    R09 & \acrfull{fs} & Past traffic remains secure upon future compromise. \\
    R10 & Cryptographic Agility & Update of cryptographic primitives enable the reaction to vulnerabilities in cryptographic systems and support \gls{pqc}. \\
    R11 & Resource Constraints & Minimal overhead in terms of computation, memory, and bandwidth to fit the protocol to the outlined use case in resource-constrained environments. \\
    R12 & Merge and Split & Taking dynamicity (R06) further, groups shall be able to split into and merge with subgroups. \\
    \bottomrule
  \end{tabular}
\end{table*}

We derive the requirements from the operational characteristics of \gls{cps} buses and our attacker model. 
In total, we outline 14 different requirements, as presented in Table \Cref{tab:req}.

In \gls{ot}, the classic CIA triad is typically reordered to AIC, prioritising availability first. 
However, availability on buses cannot be ensured solely by cryptography, but rather primarily through redundancy and system design; we therefore do not require availability properties from the GKA itself. 
By contrast, integrity and, where necessary, confidentiality are ensured by cryptographic mechanisms that utilise the group key established by the \gls{gka}. 
Given that many bus messages are short-lived and broadcast, confidentiality is often secondary, while integrity and origin assurance of messages are central. 
This leads directly to authenticity (R01): only authenticated participants may send valid protocol messages.
To avoid single points of failure and support deployments without dependable infrastructure, the protocol must be fully distributed (R02) and operate independently of external services during normal use (R03). 
Because industrial deployments vary widely in physical and link-layer properties, the protocol should be bus-agnostic (R04), i.e., not rely on bus-specific addressing or ordering features; we only assume messages are large enough to carry required cryptographic material (excluding only protocols with very small payloads such as classic 8-byte \gls{can}).
Industrial systems might experience membership changes at runtime (power-cycling by operators, hot-swapping, or unexpected outages). 
Hence, dynamic joins and leaves are required (R05), including the removal of inactive nodes without disrupting normal operation. Modern buses can host 100+ devices; therefore, the protocol must scale in state, computation, and messaging (R06).
Our adversary follows the Dolev–Yao model with full control of the channel, including eavesdropping, injection, modification, replay, and dropping. 
The protocol must remain robust under such conditions (R07). 
While a Dolev–Yao adversary can always degrade availability (e.g., by partitioning or suppressing traffic), if availability is not targeted, we aim for the strongest feasible guarantees for integrity and authenticity.
Compromise resilience is essential over long device lifetimes. 
We therefore require \gls{pcs} (R08) so that the system can 'heal' after a temporary state exposure, and \gls{fs} (R09) so that past traffic remains secure even if long-term keys are later compromised. 
Cryptographic agility (R10) is necessary to respond to evolving cryptanalysis and to facilitate \gls{pqc} adoption over a multi-decade lifecycle. 
At the same time, \gls{cps} devices are resource-constrained in CPU, memory, and bandwidth; the protocol must minimise overhead (R11) and remain feasible on constrained buses.
Finally, some \gls{cps} deployments benefit from securely reshaping groups in the field (e.g., trains coupling/decoupling). 
Supporting merge and split of groups (R12) is desirable to extend dynamicity beyond single-node joins/leaves and to better match real operational workflows.

\section{Analysis of GKA Protocols} \label{sec:protocol_analysis}

To find a suitable candidate for the \gls{gka}, we analysed different protocols. These different protocols are evaluated in light of the requirements outlined in Section \ref{sec:req}. Our analysis is oriented along the taxonomy of the works of Prandtl et al. \cite{prantl_survey_2022}. Additionally, our analysis also includes the taxonomy by Xiong et al. \cite{xiong_survey_2019}, which covers several aspects of our requirements. 


For our purposes, we exclude the following classes of group key agreements to find a suitable candidate for our use case. 
According to the study by Prandtl et al. \cite{prantl_survey_2022}, we exclude the classes of centralized and decentralized approaches, as these classes are in direct conflict with requirement R02 Fully Distributed. 
From Xiong et al.'s \cite{xiong_survey_2019} work, we exclude imbalanced \gls{gka} protocols and cluster-based protocols for the same reason. 
While a class of GKA protocols exists that generate a shared asymmetric key, this represents only a minority of protocols. 
To prevent confusion, we will treat all GKAs as if they exclusively generate symmetric keys, as this is the case for the vast majority of protocols. \cite{xiong_survey_2019} 
Those asymmetric \gls{gka} protocols would indirectly violate R11 resource constraints, since they would require computationally more expensive operations when compared with symmetric keys. 
Any protocol requires to be dynamic, see requirement R05, which excludes any protocol from the other works \cite{xiong_survey_2019, prantl_survey_2022} that do not support dynamic changes of the group.

Next to those properties, the future security guarantees are also important, as outlined by the requirements R08-10. 
After exclusion of protocols that do not sufficiently cover R08-10, a list of suitable protocols has been identified. \Cref{tab:comp_performance} shows the protocols and lists their theoretical complexity to estimate the degree of fulfillment of R11 resource constraints.

\begin{table}[htb!]
    \centering
    \caption{Performance comparision of candidates for \gls{gka} on \gls{cps} buses}
    \begin{tabularx}{\columnwidth}{|X|c|c|}
        \hline
        Protocol & Computional Overhead & Message Overhead \\
        \hline
        TGDH \cite{kim_Treebased_2004} & $O(log_2(n))$ & $O(log_2(n))$ \\
        \hline
        D-OFT \cite{dondeti_Distributed_1999} & $O(log_2(n))$ & $O(log_2(n))$ \\
        \hline
        SGRS \cite{bilal_Secure_2017} & $O(n)$ & $O(n)$ \\
        \hline
        GDH.2 \cite{steiner_DiffieHellman_1996} & $O(n)$ & $O(n)$ \\
        \hline
        GDH.3 \cite{steiner_DiffieHellman_1996} & $O(n)$ & $O(n)$ \\
        \hline
        Dutta \& Barua \cite{dutta_Constant_2005} & $O(n)$ & $O(n)$ \\
        \hline
        Kim et al. \cite{kim_ConstantRound_2004} & $O(n)$ & $O(n)$ \\
        \hline
        TreeKEM \cite{bhargavan_TreeKEM_2018} & $O(log_2(n))$ & $O(log_2(n))$ \\
        \hline
    \end{tabularx}
    \label{tab:comp_performance}
\end{table}

The performance comparison shows that the protocols TGDH \cite{kim_Treebased_2004}, D-OFT \cite{dondeti_Distributed_1999}, and TreeKEM \cite{bhargavan_TreeKEM_2018} are optimal in terms of overhead. 
They yield the same results for both computational and message overhead. 
Due to the production readiness of TreeKEM, as specified as the core protocol of \gls{mls} \cite{barnes_Messaging_2023}, we have chosen to use TreeKEM as the baseline for our \gls{gracybus} protocol. 
Using \gls{mls} directly is not possible as it requires an authentication and delivery service, which both would violate multiple requirements, in particular R02 Distributed, and R03 Independent. This adoption of the protocol gives rise to our protocol \gls{gracybus}.



TreeKEM \cite{bhargavan_TreeKEM_2018}
protocol to establish a key shared by a group of members, each controlling multiple devices who shall be able to participate asynchronously and can issue asynchronous group modification requests (add new
members, remove members, update own keys, etc.)

LL-CGKA \cite{dodis_EndtoEnd_2023}
designed as a 'leader-based continuous group key agreement with liveness'

Through the process of deduction, we have identified TreeKEM as the most suitable protocol for our use case. 
Unfortunately, we cannot simply use an \gls{mls} implementation of TreeKEM. 
This is because \gls{mls} requires an Authentication Service (AS) and a Delivery Service (DS). 
Due to the distributed nature of our use case, as well as requirements R02 (No Central Point of Failure) and R03 (Independent Operation), \gls{mls} is unsuitable for our use case. 
While \gls{mls} is not suitable for the use case, inspiration can be taken from the TreeKEM protocol and its implementations.
It should be noted that the TreeKEM proposal \cite{bhargavan_TreeKEM_2018} only provides a high-level definition of how the protocol works. 
The exact messages that must be transferred are not listed. 
Additionally, TreeKEM does not support any form of authenticity guarantees. 
Hence, TreeKEM must be modified to fit the use case. 
The resulting protocol and the messages it requires will be described in Section \ref{sec:gracy}.

\section{GRACYBUS Protocol Design} \label{sec:protocol_design} \label{sec:gracy}
Based on our analysis, the TreeKEM protocol, part of the \gls{mls} standard, provides a well-suited foundation.
However, it needs modifications and additional features to function on a \gls{cps} bus. 
We name the resulting protocol \gls{gracybus}.
This section introduces the protocol’s core elements, such as the tree structure, epoch key schedule, and bus-appropriate messaging.

\subsection{Design Overview} 
\gls{gracybus} is a fully distributed, authenticated group key agreement designed for broadcast, half-duplex \gls{cps} buses. 
It builds on TreeKEM’s contributory, tree-based design and adapts it to bus realities by removing MLS’s centralized services, minimizing asymmetric work, and leveraging broadcast encryption to subtrees. 
The protocol meets the derived requirements (R01–R12) by combining a perfect binary key tree, an epoch key schedule, a minimal set of operations (UPDATE, JOIN, LEAVE) \cite{steiner_Key_2000, steiner_CLIQUES_1998}, PKI‑based authenticity bootstrapping, and bus‑centric messaging with replay protection and robustness features.

\subsection{GRACYBUS Tree Structure}
\gls{gracybus} organises members in a perfect binary tree. 
Each leaf corresponds to one device; non‑leaf secrets are derived as hashes of the last‑updated child secret.
Every device holds private keys for nodes on its path and public keys for nodes on its co‑path. 
To avoid stalls in third‑party LEAVE, \gls{gracybus} further stores all active leaf public keys (redundancy), trading O(n) public‑key storage for robustness (R03, R11).
The tree grows via leftmost‑unused insertion on JOIN;
when the right subtree becomes empty, truncation reduces height (simplifying updates and improving efficiency).

\subsection{Epochs and Key Schedule}

Any UPDATE, JOIN, or LEAVE creates a new epoch. The group traffic key is derived by a KDF from the current root secret and the previous epoch key:
\begin{equation}
\label{eq:epoch_key}
EpochKey_e = KDF(Secret_e, EpochKey_{e-1})    
\end{equation}

Past epoch keys are erased and not retained by participants; messages are assumed to be short‑lived (R09). 
This, combined with the overwriting of path keys, supports \gls{fs} and \gls{pcs} (R08, R11).

\subsection{Identity and Authenticity Bootstrapping}
Identity and authenticity rely on \gls{pki}: devices present a certificate (rooted in an industrial trust anchor) and sign JOIN handshake messages with their certificate private key. 
Within the group, integrity/authenticity uses \glspl{mac} from the current epoch key (R01, R03). 
Revocation infrastructure is out of scope; operational guidance is to protect certificate private keys

\subsection{GRACYBUS Operations: UPDATE, JOIN, LEAVE}
UPDATE: A member rotates its leaf secret, propagates hashed secrets up its path, and broadcasts updated public keys. 
Subtree secrets are encrypted to co‑path public keys; the message is \gls{mac}‑protected with the current epoch key. 
Complexity and message size scale with tree height $\mathcal O(\log (n))$.

JOIN: An authenticated, signed handshake between joiner and sponsor bootstraps identity and transfers the joiner’s preliminary public key and hashed secret. 
The sponsor inserts the leaf, computes affected secrets, and: (1) sends updated secrets/public keys to current members (MAC‑protected), and (2) delivers the new epoch key to the joiner (encrypted to the joiner’s certificate key). 
The sponsor also shares all active leaf public keys with the joiner to enable robust LEAVE later. 
Sponsor selection prefers the rightmost feasible leaf; expansion cases are handled by any member.

LEAVE: Executed by another member. 
New secrets are set on the leaver’s path and are encrypted only to public keys that the leaver cannot decrypt. 
Voluntary leaves use a MAC-protected Leave Request; unannounced leaves are detected by an update window. 
Members must periodically update; otherwise, 
LEAVE messages mirror UPDATE (MAC‑protected), with encryption to appropriate co‑path keys. Truncation applies when the right subtree becomes empty.

\subsection{Messaging: Formats and Processing}
Messages are structured for broadcast environments: secrets are encrypted with the public keys of subtrees, allowing multiple recipients to decrypt a single broadcast. Public keys are shared as plaintext (integrity-protected), while secrets are shared as ciphertext (confidentiality). 
UPDATE and LEAVE carry PublishPublicKey and UpdateNodesSecretKey sub‑messages and are MAC‑bound to the current epoch (replay‑safe). 
JOIN handshake uses signatures and nonces to prevent forgery and replay. 
\gls{gracybus} assumes a total-order reliable broadcast on the bus for delivery and ordering; segmentation/fragmentation are handled below the protocol (R04). 

\subsection{Cryptographic Agility and Cipher Suites}
Operations scale with tree height $\mathcal O(\log (n))$; only the joiner’s receipt of all leaf public keys scales with the current group size (robustness trade‑off). Storage remains lightweight for private keys (O(log n)), with O(n) public keys for redundancy. 
All crypto interfaces are algorithm‑agnostic (hash, KDF, MAC, signature, encryption/KEM), enabling cryptographic agility and PQC adoption over long lifecycles (R10, R13, R11).



\section{Evaluation} \label{sec:evaluation}

The developed protocol enables the efficient computation of the shared secret through its tree structure, which limits any group operation to the height of the tree, as discussed in Section \ref{sec:gracy}. Therefore, any group operation requires $\mathcal O(\log_2(n))$ operations. Further discussion of the computational overhead can be found in \ref{ann:comp}.

Since \gls{cps} bus systems are not only limited in their computational resources, but also in available bandwidth, we also evaluate the required messages for our protocol. Table \ref{tab:eval_performance} compares the different messages corresponding to the group operations. Some messages, such as Join Requests, are of a constant size, meaning they are not dependent on the size of the key tree. Other messages require $\mathcal O(\log_2(n))$ messages to inform participants about an update to the key tree. A special case is the Join Failed message, which is triggered based on the number of Join Send Secret messages that need to be rejected. Therefore, this message in particular is not bound by the key tree itself.
\begin{table}[!htb]
    \centering
    \caption{Comparision of \gls{gracybus} Message Sizes}
    \begin{tabularx}{\columnwidth}{|X|c|c|}
        \hline
        Message & Constant Length & Bounded by \\
        \hline
        Join Request & $\checkmark$ & \\
        \hline
        Join Challenge & $\checkmark$ & \\
        \hline
        Join Send Secret & $\checkmark$ & \\
        \hline
        Join Success (Joining Node) & $\times$ & $\log_2(n)$\\
        \hline
        Join Success (GKA) & $\times$ & $\log_2(n)$\\
        \hline
        Join Success Combined & $\times$ & $\log_2(n)$\\
        \hline
        Join Failed & $\times$ & $\infty$ \\
        \hline
        Update & $\times$ & $\log_2(n)$ \\
        \hline
        Leave Request & $\checkmark$ & \\
        \hline
        Leave Update & $\times$ & $\log_2(n)$\\
        \hline
    \end{tabularx}
    \label{tab:eval_performance}
\end{table}

Not only is bandwidth on buses limited, but the systems operating on those are also suffering from very constrained resources. We therefore provide an overview of the required storage requirements for \gls{gracybus}. Table \ref{tab:eval_storage} gives an overview of the storage requirements for the different key types of the key tree.

\begin{table}[htb!]
    \centering
    \caption{Storage Requirements of \gls{gracybus}}
    \begin{tabularx}{\columnwidth}{|X|X|p{3.25cm}|}
        \hline
        Key type & Number of keys & Sources \\
        \hline
        Public keys & $2n+1$ & \begin{itemize}\item $2n-1$ key per node in tree
            \item $1$ from certificate
            \item $1$ from \gls{ca}\end{itemize}\\
        \hline
        Private keys & $\lceil \log_2(n)\rceil+1$ & 
        \begin{itemize}
            \item $\lceil\log_2(n)\rceil$ key for nodes on path
            \item $1$ from certificate
        \end{itemize}\\
        \hline
        Epoch keys & $1$ & \begin{itemize}
            \item $1$ from current epoch
        \end{itemize}\\
        \hline
    \end{tabularx}
    \label{tab:eval_storage}
\end{table}
To summarize, our protocol supports efficient group key agreement with a logarithmic upper bound on computational resources, the required bandwidth through message sizes, and the storage requirements for the different keys. Therefore, our protocol satisfies the requirement R11 and can be used in resource-constrained environments.

We evaluate the security of the proposed protocol with respect to the threat model outlined in Section \ref{sec:att_model}. Table \ref{tab:eval_crypt} provides an overview of the various protection mechanisms employed to prevent eavesdropping on sensitive data, ensure the integrity of join attempts, and thereby safeguard the integrity of group communication and its shared secret. The exact cryptographic primitives used are outlined in Section \ref{sec:gracy}. 

Since the Dolev-Yao attack is unable to break cryptographic primitives, our protocol ensures group integrity through multiple mechanisms. All group joins are protected using signatures, allowing only authenticated nodes to join the group. Sensitive information, such as secrets, is encrypted to prevent the leakage of key material. All final group update messages that change the state of the key tree are secured using a \gls{mac}, ensuring the consistency of the key tree. A Dolev-Yao attacker can still partition the bus into multiple parts, which would yield different key trees in the corresponding partitions, effectively compromising the system's overall availability. Therefore, we only partially satisfy the requirement R07. Note that this weakness comes from the attacker's capability to drop messages. Any other key agreement protocol would suffer from the identical partitioning problem. 
\begin{table}[htb!]
    \centering
    \caption{Protection Mechanisms for Group Messages}
    \begin{tabularx}{\columnwidth}{|p{3cm}|X|X|X|}
        \hline
        Message & Encrypted & MAC & Signature \\
        \hline
        Join Request & $\times$ & $\times$ & $\times$ \\
        \hline
        Join Challenge & $\times$ & $\times$ & $\checkmark$ \\
        \hline
        Join Send Secret & $\checkmark$ & $\times$ & $\checkmark$ \\
        \hline
        Join Success & $\checkmark$ & $\checkmark$ & $\checkmark$\\
        \hline
        Join Failed & $\times$ & $\times$ & $\checkmark$ \\
        \hline
        Update & $\checkmark$ & $\checkmark$ & $\times$ \\
        \hline
        Leave Request & $\times$ & $\checkmark$ & $\times$ \\
        \hline
        Leave Update & $\checkmark$ & $\checkmark$ & $\times$ \\
        \hline
    \end{tabularx}
    \label{tab:eval_crypt}
\end{table}

We evaluate the capability of our protocol to be robust against compromise with the properties of \gls{pcs} and \gls{fs}. Also, the ability to protect against future attackers with quantum computing capabilities is evaluated. Those align with the requirement R08-10 in Section \ref{sec:req}.

Leaked key material is a significant problem for most cryptographic algorithms, as it enables an attacker to decrypt all traffic. To prevent this, we evaluate the properties of \gls{pcs}, which means that if the key material is leaked at some point in time, future communication can still be secured. This assumes the attacker has no access to the secret authentication keys, since otherwise healing is not possible. These assumptions correspond to the adversary model with \gls{pcs} through limited compromise as outlined in \cite{cohn-gordon_PostCompromise_2016}. Under the assumption of permanent full compromise, \gls{pcs} cannot be achieved realistically because of the possibility of state loss, as already outlined in \cite{11023469}. Nevertheless, if an update of the key material is successful without the adversary's interference, our protocol heals, as the adversary is unable to generate the new epoch key without the hashed secret.

\gls{fs} describes the ability that an attacker cannot decrypt past traffic, even in the case that the long-term keys of all devices have been leaked. As shown in eq. \ref{eq:epoch_key}, a \gls{kdf} for the generation of epoch keys is used. This KDF uses an ephemeral secret and the old epoch key as input; therefore, an attacker cannot efficiently calculate past epoch keys, even after the system is compromised. Our protocol fulfils the requirement of \gls{fs} through this design decision.

In the near future, an attacker with quantum capabilities may become a possibility. To prepare for such an event, we assess the security capabilities of our protocol in the event of a compromise. This affects all messages that require a signature for message authenticity, particularly those involving a quantum computer, which lowers the security level of asymmetric cryptography based on the discrete logarithm problem \cite{shor_PolynomialTime_1999}. By design, \gls{gracybus} protects the messages without specifying the exact cryptographic primitives, keeping the protocol cryptographically agile. We provide the cryptographic primitives used for the implementation in Appendix \ref{ann:impl}. We want to emphasise that the security level of cryptographic primitives can change at any time. 

Our protocol \gls{gracybus} is therefore robust against compromise of key material and against an attacker with quantum capabilities. Therefore, it satisfies the requirements R08, R09, and R10.

\section{Conclusion} \label{sec:conclusion}

\acrlong{cps} that relying on shared communication buses, demand protection mechanisms that do not compromise the stringent performance requirements inherent to these environments. Achieving meaningful security in such systems requires approaches that can efficiently generate and distribute shared secrets while adhering to modern cryptographic standards.

The protocol introduced in this work demonstrates that it is feasible to design key-establishment mechanisms tailored to bus-oriented CPS settings without weakening security guarantees. By explicit design \gls{gracybus} combines constraints on computation and communication with group key establishment. More importantly, it integrates contemporary security principles, such as \gls{fs} and \gls{pcs}, as well as readiness for post-quantum adversaries—capabilities that are traditionally difficult to realise in multi-party, resource-restricted systems.

This contribution outlines future-ready \gls{pcs} protection strategies for bus systems. By aligning efficient group key generation with rigorous, state-of-the-art cryptography, we show that high-assurance security can be embedded directly into \gls{cps} infrastructures. With an efficient key establishment method, our approach paves the way for protecting \gls{cps} communication via encryption and integrity protection, leveraging modern underlying security principles.




Several further improvements of the protocol design are possible and should be addressed by future work. One example of such advancements is the requirements outlined in R12 in Section \ref{sec:req}. The designed protocol does not include any group operations, such as merging or splitting, of different groups. Those operations also need to be designed securely for past and future session keys. After a merge, the groups should not have access to prior epoch keys, while after a split, the groups should not have access to the epoch key of the other one. 

In addition to the merge and split operations, it would also be possible to batch or parallelise operations together, allowing for more efficient utilisation of the key tree structure by avoiding empty leaves. An example would be a combination of leave and join operations. Those parallelisations also would have implications for protocol security and need to be analysed thoroughly. 

Another optimisation of the protocol would be to improve the truncation or reordering of the tree in order to delete unused empty leaves. Currently, the protocol design only allows truncation if the right subtree has empty leaves. This behaviour can lead to linear complexity in the worst-case scenario, particularly in terms of the number of active participants. 

%
%
%

\section*{Acknowledgements}
This work was supported by the Federal Ministry for Economic Affairs and Energy of Germany, which funded the project ATLAS-L4 within the research program 'Neue Fahrzeug- und Systemtechnologien'.

\bibliographystyle{unsrt}
\bibliography{agrok}

\appendices

\section{Apendix}

\subsection{Computational Overhead per Group Operation}
\label{ann:comp}
To analyze the protocol with respect to performance overhead, we list the number of operations required to successfully perform a specific group operation. The designed protocol is cryptographically agile, meaning that, based on the specific cryptographic operation and the underlying hardware, the actual overhead can vary. Therefore, we analyze the protocol in an abstract manner, based on the required operation, without considering the specific cryptographic algorithm. We introduce the following terminology for elemental operations.

\begin{table*}[ht]
    \centering
    \caption{Protection Mechanisms for Group Messages}
    \begin{tabularx}{\textwidth}{|p{3.5cm}|p{4.5cm}|X|X|}
        \hline
        \textbf{Message} & \textbf{Generation} & \textbf{Verification} & \textbf{Update} \\
        \hline
        Join Request & - & - & - \\
        \hline
        Join Challenge & $G+S$ & $2V$ & - \\
        \hline
        Join Send Secret & $2G+H+K+S$ & $2V+N$ & - \\
        \hline
        Join Success (Joining Node) & $D+E+S$ & $V+N$ & $(h-2)*(H+K)+E$\\
        \hline
        Join Success (Other Nodes) & $(h-2)*(H+K)+(h-1)*K+M$ & $M$ & $E+(h-2)*H+(h-1)*K$\\
        \hline
        Join Failed & $S$ & $V$ & - \\
        \hline
        Update & $G+(h-1)*(H+E+K)+M$ & $M$ & $E+(h-2)*H+(h-1)*K$ \\
        \hline
        Leave Request & $M$ & $M$ & - \\
        \hline
        Leave Update & $G+(h-1)*(H+E)+(h-2)*K+M$ & $M$ & $E+(h-2)*H+(h-1)*K$ \\
        \hline
    \end{tabularx}
    \label{app:tab:costs}
\end{table*}

\begin{itemize}
    \item $E$: Encryption of data using a public key
    \item $S$: Signing a messsage
    \item $M$: Generating a \gls{mac}
    \item $G$: Generating a random value
    \item $V$: Verifying a signature
    \item $K$: Generate an asymmetric key pair from a shared secret (used as a random seed)
    \item $H$: Hashing a value
    \item $D$: Derviation of a key via \gls{kdf}
    \item $N$: Verifying equality of two nonces
    \item $h$: Not an operation, but describes the height of the key tree with $h=\lceil\log_2(n)\rceil$, where $n$ is the number of nodes of the key tree.
\end{itemize}



We analyze the computational costs per group message and break them down into different phases. Each group operation usually consists of three phases. Phase one (Generation) involves generating the message itself. After the message is generated, it will be sent. The sending of this message does not contribute to the computational overhead and is instead covered in Section \ref{sec:evaluation}, particularly in Table \ref{tab:eval_performance}. After the message is received, verification is required, usually involving a computing operation, which is represented by phase two (Verification). The final phase (Update) involves updating the key tree on the devices, which typically entails hash operations and the use of the key derivation function. We observe the computational costs over the whole group for each operation. Table \ref{app:tab:costs} shows the costs per group operation. 

The Join Request does not have any protection-worthy content and therefore no security features. The computational overhead for any of the three phases is therefore zero. 

The Join Challenge is a part of the operation where the node to join proves that it is allowed to join by presenting a certificate. Therefore, the generation of this message requires the generation of a random value ($G$), which then will be signed using the certificate ($S$). The receiving node needs to verify this by firstly checking the signature ($V$) and then verifying that the certificate provided is allowed to join ($V$).

All operations in all phases are bound to the height of the tree, therefore showing acceptable overhead with respect to the participants of the group key agreement.

\subsection{Cryptographic Primitives used in the Implementation}
\label{ann:impl}
The group operation utilizes various cryptographic mechanisms, including signatures and encryption. Since the protocol design of \gls{gracybus} is cryptographically agile, we do not require any specific algorithms. The implementation still requires specific algorithms, which are outlined in this chapter. Table \ref{ann:tab_crypto} lists the algorithms as well as their usage in \gls{gracybus}.

Our implementation of \gls{gracybus} utilizes \texttt{AES-256} \cite{nationalinstituteofstandardsandtechnologyus_Advanced_2023} for encrypting and decrypting messages. The key encapsulation is performed using \texttt{CRYSTALS-Kyber} \cite{nationalinstituteofstandardsandtechnologyus_Modulelatticebased_2024, bos_CRYSTALS_2018}, allowing for the generation of an asymmetric key pair via a shared secret. \texttt{CRYSTALS-Kyber} is an algorithm implementing \gls{pqc} to protect from quantum attackers. For classical signatures, we use the \texttt{ecdsa\_secp521r1\_sha512} \cite{johnson_Elliptic_2001}, which can be attacked by quantum-capable attackers. As a \gls{pqc} alternative, we implement \texttt{CRYSTALS-Dilithium} \cite{ducas_CRYSTALSDilithium_2018}, which will be combined with the elliptic curve signature for a hybrid signature approach. As a hash function, we use \texttt{SHA-512} \cite{nationalinstituteofstandardsandtechnologyus_Secure_2015}, which is used as a foundation to generate \glspl{mac} with a \texttt{HMAC with SHA-512} \cite{crypto-1996-923} approach. For key derivation, we use \texttt{HKDF} \cite{krawczyk_Cryptographic_2010}, which completes the list of cryptographic algorithms needed.
\begin{table}[th]
    \centering
    \caption{Cryptographic Algorithms used for the Implementation}
    \begin{tabularx}{\columnwidth}{|p{2cm}|p{3.6cm}|X|}
        \hline
        Function & Algorithm & Library \\
        \hline
        Encryption & \texttt{AES-256} \cite{nationalinstituteofstandardsandtechnologyus_Advanced_2023} & Bouncycastle \\
        \hline
        Key Encapsulation & \texttt{CRYSTALS-Kyber} \cite{nationalinstituteofstandardsandtechnologyus_Modulelatticebased_2024, bos_CRYSTALS_2018}, & KyberKotlin \\
        \hline
        Signature & \texttt{ecdsa\_secp521r1\_sha512} \cite{johnson_Elliptic_2001} & Bouncycastle\\
        \hline
        Signature (\gls{pqc}) & \texttt{CRYSTALS-Dilithium} \cite{ducas_CRYSTALSDilithium_2018} & Bouncycastle \\
        \hline
        Hash function & \texttt{SHA-512} \cite{nationalinstituteofstandardsandtechnologyus_Secure_2015} & Java \\
        \hline
        \gls{mac} & \texttt{HMAC with SHA-512} \cite{crypto-1996-923} & Java \\
        \hline
        \gls{kdf} & \texttt{HKDF} \cite{krawczyk_Cryptographic_2010}& Bouncycastle \\
        \hline
    \end{tabularx}
    \label{ann:tab_crypto}
\end{table}
\end{document}